\documentclass{ws-rv961x669}
\usepackage{ws-rv-van}     
\usepackage{ws-rv-thm}     
\makeindex

\newcommand{\bi}{\begin{itemize}}
\newcommand{\ei}{\end{itemize}}
\newcommand{\bea}{\begin{align}}
\newcommand{\eea}{\end{align}}
\newcommand{\be}{\begin{equation}}
\newcommand{\ee}{\end{equation}}

\newcommand{\gad}{{\dot{\alpha}}}
\newcommand{\gbd}{{\dot{\beta}}}

\newcommand{\ga}{\alpha}
\newcommand{\gb}{\beta}

\newcommand{\pl}{{\partial}}

\begin{document}

\chapter[Pseudo-local Theories: A Functional Class Proposal]{Pseudo-local Theories: A Functional Class Proposal}\label{ra_ch1}

\author[Massimo Taronna]{Massimo Taronna\footnote{Postdoctoral Researcher of the Fund for Scientific Research-FNRS Belgium.}}

\address{Universit\'e Libre de Bruxelles,\\
ULB-Campus Plaine C.P. 231, B-1050 Bruxelles, Belgium\\
massimo.taronna@ulb.ac.de}

\begin{abstract}
In this article, using the language of jet space, we propose a functional class space for pseudo-local functionals. We test this functional class proposal in a number of examples ranging from string-field-theory to AdS/CFT dualities. Implications of the locality proposal at the quartic order are also discussed.
\end{abstract}


\body

\tableofcontents

\section{Introduction}\label{ra_Intro}

Locality is a fundamental concept in physics. In general terms, it is related to a classical space-time description which should emerge in the semiclassical limit of quantum gravity. Locality plays also a key role in quantum field theory and its interrelations with analyticity and unitarity at the S-matrix level have been the subject of many studies\cite{1966asm..book.....E}. In the holographic context the challenge is to reconstruct the bulk physics from the boundary CFT observables and in particular their analyticity and singularity properties\cite{Heemskerk:2009pn,ElShowk:2011ag,Fitzpatrick:2012cg,Maldacena:2015iua}. Nonetheless, from the perspective of quantum gravity in the bulk one expects the local semiclassical description to break down for energies high enough.

In the string theory context the scales at play are the Planck length $l_p$ and the string length $l_s$. So one expects to recover a semiclassical local space-time description for length scales $L$ such that\footnote{Recall the relation between Planck length, string length and string coupling constant $$g_s^2\sim\left(\frac{l_p}{l_s}\right)^{\tfrac{d-2}{2}}\,.$$} $l_p\ll l_s\ll L$. At energy scales large enough to probe the string length one can indeed excite macroscopic strings and their effect should be taken into account by the theory without breaking unitarity and causality. String scattering amplitudes for instance, predict a whole tail of corrections to the semiclassical point-particle approximation which become relevant at length scales comparable with the string-length\cite{Amati:1987wq,Gross:1987kza,Gross:1987ar,Amati:1987uf,Amati:1988tn} $L\sim l_s$. This translates in an effective pseudo-local form of string interactions which involve derivatives of arbitrary order. This is a key ingredient for the soft behaviour of high energy scattering amplitudes in string theory. Therefore, already string cubic (off-shell) vertices are pseudo-local functionals expressed as formal series of the following type:
\begin{equation}\label{pseudo}
\mathcal{V}\sim \sum_{l_1,l_2,l_3=0}^\infty c_{l_1,l_2,l_3}(\alpha^\prime\nabla)^{l_1}\Phi_1(\alpha^\prime\nabla)^{l_2}\Phi_2(\alpha^\prime\nabla)^{l_3}\Phi_3\,.
\end{equation}

To appreciate this features more closely various high energy limits are usually studied probing different kinematic regions of string scattering. One is the high-energy and fixed momentum transfer limit: $s\to\infty$ at $t$ fixed (fixed impact parameter) and the second is the $s\to\infty$ limit at $\tfrac{t}{s}$ fixed (fixed angle scattering). The main question to address in this limits is to understand which among string effects or gravitational effects dominate. Remarkably both limits show asymptotic behaviours which are quite far from those of a local field theory. The $s\to\infty$ at fixed $t$ behaviour of the graviton amplitude is given (in $\alpha^\prime=\tfrac12$ units) by\cite{Amati:1987uf,Amati:1987wq,Amati:1988tn}:
\begin{equation}
\mathcal{A}\sim g_s^2\,\frac{\Gamma(-\tfrac{t}{8})}{\Gamma(1+\tfrac{t}8)}\,s^{2+\frac{t}{4}}e^{-\tfrac{t}{4}}\,,
\end{equation}
which should be compared with the pure gravity result $\mathcal{A}\sim \tfrac{s^2}{t}$ and differs from it by an effective form factor. The $s\to\infty$ at fixed angle scattering, which is a truly short distance limit of the theory due to $t\to\infty$, gives similarly:
\begin{equation}
\mathcal{A}\sim \frac{g_s^2}{(stu)^3}\,e^{-\tfrac14(s\log s+t\log t+u\log u)}\,.
\end{equation}
The exponentially suppressed behaviour above was argued to be related to a violation of polynomial boundedness of the amplitude\cite{Gross:1987kza}.
At length scales $L\sim l_s$ or $L< l_s$ one thus expects a break-down of usual effective-field theory locality, as the whole infinite higher-derivative tail \eqref{pseudo} present in the interactions becomes relevant\cite{Moeller:2005ez,Giddings:2007bw,Taronna:2010qq,Sagnotti:2010at,Fotopoulos:2010ay,Polyakov:2010sk,Polyakov:2015usr}. Locality may then be reconsidered or generalised. Most importantly a key question that should be addressed in this context, if one insists in a higher-spin theory description of string theory, is which pseudo-local field redefinitions are admissible, if any. In other words, we should be able to specify a functional class of admissible pseudo-local functionals.

This questions are even more relevant in the tensionless limit of string theory: $l_s\sim\sqrt{\alpha^\prime}\to\infty$. Higher-spin theories are conjectured to effectively describe such limits of quantum gravity\cite{Sezgin:2002rt,Klebanov:2002ja} in AdS space-time and they turn out to be naturally pseudo-local\cite{Boulanger:2008tg,Kessel:2015kna,Boulanger:2015ova}. The higher-spin algebra mixes indeed derivatives of arbitrary orders into higher-spin multiplets and requires infinitely many HS fields as well. HS vertex themselves have a number of derivatives bounded from below by the highest spin involved into the interaction\cite{Metsaev:1991mt,Metsaev:2005ar,Boulanger:2008tg,Vasilev:2011xf,Joung:2011ww,Boulanger:2012dx,Boulanger:2013zza}. On top of this one should carefully take into account the infinite sums over spins that accompany and infinite higher-derivative contributions. There should be in fact a close relation between higher-spin exchanges in the direct channel and higher-derivatives contributions in cross channels. Crossing requires the latter to reconstruct the former and vice-versa. This may also play a key role in defining a tensionless limit in string theory where all poles in Virasoro amplitude degenerate to a single massless pole\footnote{Some recent results on conformal higher-spin theories around flat space appeared recently\cite{Joung:2015eny}.}. Similar issues may as well arise in curved backgrounds but the status in this case is not as clear due to the luck of an handy formulation of string theory.

Here, we will try and address this questions in simple examples ranging from bosonic string cubic interactions to conserved currents in AdS space. We also analyse some of the consequences of our discussion at the quartic order and in the AdS/CFT context making use of Mellin amplitudes techniques.

This article is organised as follows. In Section \ref{sec:pseudo}, we introduce the main concepts to deal with locality in a field theory framework. In particular, we introduce the concept of jet space and inverse limit to define pseudo-local interactions. In Section~\ref{sec:Examples} we apply the logic discussed in Section~\ref{sec:pseudo} to various examples ranging from string theory to conserved currents in $AdS_3$ and $AdS_4$. In Section~\ref{sec:CFT} we summarise some results available on the CFT side and discuss locality in this framework. The conclusions are summarised in Section~\ref{sec:Conclusions}.

\section{Pseudo-Locality and Inverse Limit}\label{sec:pseudo}
The goal of this paper is to study a framework that would allow to treat the problem of locality in the context of field theories with infinitely many derivatives. This is a difficult problem and here we will summarise and push forward some ideas recently appeared in the literature\cite{Vasiliev:2015wma,Skvortsov:2015lja}. For definiteness, we shall be interested in the analysis of Noether current interactions at the equations of motion level. We shall comment on the extensions and implications of this analysis to higher orders in some examples. In this section we introduce the main ideas.

\paragraph{Jet space:} The standard tool to address locality in field theory is jet space. Given a set of fields $\phi^I(x)$, which are usually tensors on some tensor bundle $\pi:E\rightarrow M$, the central idea is to promote them, together with all of their derivatives of arbitrary order, to new independent coordinates. Using the notation:
\be
[\phi(x)]^k=\left\{C^I(x),C_\mu^I(x),\ldots,C_{\mu(k)}^I(x)\right\}= \left\{\phi^I(x),\pl_\mu\phi^I(x),\ldots,\pl_{\mu(k)}\phi^I(x)\right\}\,,
\ee
one first defines the space $V^k$ of local functionals of order-$k$. This is the space whose coordinates are given by $[\phi]^k$:
\be
V^k=\left\{[\phi(x)]_k\ \big|\ \phi(x):M\rightarrow E\right\}\,.
\ee
Then, one defines the jet space of order $k$ as $J^k=M\times V^k$. The jet space of order zero coincides with the original fibre bundle $E$. Each section on $E$ induces a section over $J^k(E)$ through the identification $C^I_{\mu(k)}\sim\pl^k\phi^I$. We thus have a sequence $J^k$ of topological spaces. Furthermore, the projection maps $\pi_{k,l}$ ($k>l$) defined as:
\be
\pi_{k,l}([\phi]_k)=[\phi]_l\,,
\ee
are smooth fibre bundles.

A sequence of topological spaces as above endowed with maps $\pi_{k,k-1}$:
\be
\cdots\longrightarrow J^{k+1}\xrightarrow{\pi_{k+1,k}}J^k\longrightarrow\cdots\longrightarrow J^1\xrightarrow{\pi_{1,0}} E\xrightarrow{\pi}M\,,
\ee
such that:
\be
\pi_{i,j}\circ\pi_{j,k}=\pi_{i,k}\,,\quad\forall i>j>k\,,
\ee
is called in mathematical terms an \emph{inverse sequence} and it defines the infinite jet space $J^\infty$ as the following commutative diagram\cite{Krasil'shchik:2010ij}:
\be
\begin{matrix}\includegraphics[width=6cm,keepaspectratio]{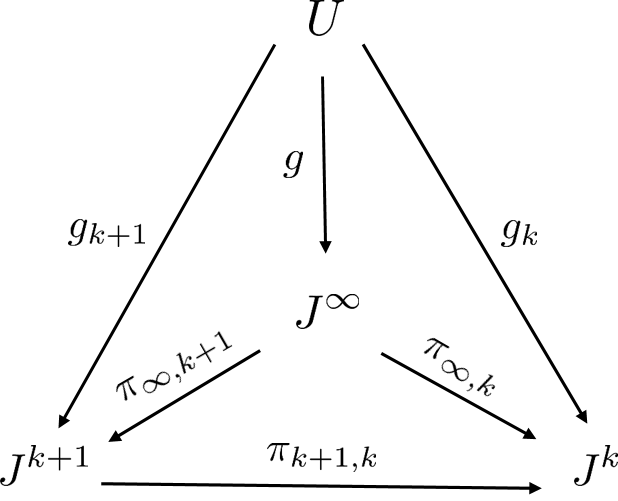}\end{matrix}\,.
\ee
The above is \emph{unique} and \emph{universal} when it exists.

For later convenience it is also useful to introduce the space of quadratic functions $Q_k$ over $J^\infty$ with at most $k$ derivatives. This will be the natural space where to study Noether currents. The space $Q_k$ is simply obtained by decomposing the tensor products of two jets and considering the corresponding linear vector space. A choice of basis for $Q_k$ is for instance:
\be
[\phi\times \phi]_k=\left\{C^I_{\mu(m)}(x)C^{J}_{\nu(n)}(x)\right\}_{m+n\leq k}\,,
\ee
which allows again to define continuous $\pi_{k,l}$ maps as:
\be
\pi_{k,l}([\phi\times \phi]_k)=[\phi\times \phi]_l\,.
\ee
One can then proceed to define the limiting space $Q_{\infty}$ of quadratic functionals of infinite order in derivatives. The space $Q_\infty$ is the main object we aim to study in this article. Elements $j\in Q_{\infty}$ can be defined starting from the following formal series:
\be
\tilde{j}_{\mu(s)}=\sum_{n,m}(f_{JK})_{\mu(s)}{}^{\nu(n),\rho(m)}\,C^J_{\nu(m)}(x)C^{K}_{\rho(n)}(x)\,,\label{Qelem}
\ee
where the tilde $\tilde{j}$ notation is meant to stress that the corresponding object one starts with is a formal series. Here, $(f_{JK})_{\mu(s)}{}^{\nu(n),\rho(m)}$ are structure constants parametrising the decomposition of the tensor product of two jets.
The maps $\pi_{\infty,k}$ are then defined as:
\be
\pi_{\infty,k}(\tilde{j}_{\mu(s)})=\sum_{m+n\leq k}(f_{JK})_{\mu(s)}{}^{\nu(n),\rho(m)}\,C^J_{\nu(m)}(x)C^{K}_{\rho(n)}(x)\,,
\ee
so that the actual definition of $j\in Q_{\infty}$, from \eqref{Qelem}, is the following limit:
\be
j_{\mu(s)}= \lim_{k\to\infty}\pi_{\infty,k}(\tilde{j}_{\mu(s)})\,.
\ee
If the above limit exists the inverse-limit construction guarantees its uniqueness.

So far we have introduced the off-shell jet space space $J^\infty$ on which we have defined quadratic functionals $Q_\infty$. One can also define an \emph{on-shell} jet space often called in the literature \emph{stationary surface} $\Sigma^\infty\subset J^\infty$. Given some equations of motion $L^I([\phi])=0$, all of their higher-derivative consequences $[L^I([\phi])]=0$ define an algebraic equation on $J^\infty$. One first defines:
\be
\Sigma^k=\left\{[\phi]^k\in J^k\,\big|\,[L^I([\phi])]^k=0\right\}\subset J^k\,,\label{stationary}
\ee
as the subspace of the order-$k$ jet $J^k$ defined by the equations $[L^I([\phi])]^k=0$. One then take the limit $k\to\infty$.
In terms of the tensorial coordinates which define the off-shell jet, the stationary surface becomes an algebraic condition which amounts to some irreducibility projection for the original tensorial coordinates. For instance, in the case of a scalar on Minkowski space, the irreducibility condition simply amounts to a tracelessness condition:
\be
[\phi(x)]^k\approx\left\{C(x),C_\mu(x),\ldots,C_{\mu(k)}(x)\right\}\,,\qquad C^{\nu}{}_{\nu\mu(k-2)}=0\,.\label{stationary2}
\ee
In the presence of gauge symmetries it is also important to take into account that not all equations of motion $L^I([\phi])$ are independent. As a consequence of gauge symmetries the equations should satisfy some relations which are called Noether or Bianchi identities:
\be
[\mathcal{B}^I([L])]=0\,.
\ee
They parametrise all redundancies among the equations $L^I([\phi])$ themselves and all higher-derivative consequences thereof $[L^I([\phi])]$. In the following we shall use the week equality symbol $\approx$ whenever an equality is assumed to hold on the stationary surface. Furthermore, given the stationary surface defined by the \emph{linear} equations of motion: ${\Sigma}^\infty$, one can consider the on-shell space $\bar{Q}_\infty$ of quadratic functionals defined on such stationary surface $\bar{Q}_\infty\approx Q_\infty$.

\paragraph{Unfolding and Jet Space}

Unfolded equations acquire a natural interpretation from the point of view of jet space \cite{Barnich:2004cr,Barnich:2010sw,Grigoriev:2012xg,Rahman:2015pzl}. One can indeed consider a decomposition of $J^\infty$ into modules of the rigid symmetry algebra of the background. Geometry of the jet bundle is then encoded into appropriate nilpotent form-degree one covariant derivatives ${\mathcal{D}}$, ${\mathcal{D}}^2=0$. The linear dynamical equations take the form of covariant constancy conditions:
\be
{\mathcal{D}}C^{\mathsf{A}}=(d-\omega^{ab}\hat{L}_{ab}-h^a\hat{P}_a)C^\mathsf{A}=0\,,
\ee
directly in terms of the stationary surface variables \eqref{stationary}.
For ease of notation, and to be more uniform with the literature on the unfolded formalism, we have renamed the jet variables as $C^\mathsf{A}=[\phi]$.

In the unfolded language one identifies the stationary surface coordinates at the linear level with $0$-forms, owing to their covariance under rigid transformations. On the other hand, the gauge dependent components can be introduced as connections gauging the (HS-)global symmetries. This generalise the spin-connection $\omega^{ab}$ and the vielbein $h^a$ in the gravity case \cite{Lopatin:1987hz,Vasiliev:1988sa,Vasiliev:1999ba}. HS connections $\omega^{\mathcal{A}}$ transform in a different module of the isometry group and their gauge dependence is naturally encoded into the following gauge transformations:
\be
\delta \omega^{\mathcal{A}}=\mathcal{D}\epsilon^{\mathcal{A}}\,.
\ee
Above $\epsilon^{\mathcal{A}}$ is a gauge-symmetry parameter. The module in which it transforms defines the adjoint module of the HS algebra.

To summarise, the unfolding formalism relies on a reformulation of the dynamics directly in terms of the stationary surface coordinates. These are decomposed in gauge-dependent and gauge-covariant components transforming as modules under the (HS) rigid symmetries of the theory. In the following we will mostly concentrate our attention on gauge-invariant current interactions which are build out of $0$-forms only.

\paragraph{The ``current'' interaction problem:}
At the equations of motion level we consider HS-currents $j_{\mu(s)}(C,C)\in \bar{Q}_\infty$ as quadratic sources to linear HS equations:
\be
\mathcal{F}_{\mu(s)}(\phi)\equiv(\nabla^\mu\nabla_\mu-m^2)\phi_{\mu(s)}+\ldots=j_{\mu(s)}(C,C)\,.
\ee
Depending on the eventual Bianchi identities\footnote{For instance the Bianchi identity satisfied by the Fronsdal operator on AdS is:
\be
\nabla^\mu\left(\mathcal{F}_{\mu(s)}-\frac12 g_{\mu(2)}\mathcal{F}^\prime_{\mu(s-2)}\right)=0\,.
\ee}:
\be
\mathcal{B}(\mathcal{F}_{\mu(s)})=0\,,
\ee
satisfied by the kinetic operator, the current should also satisfy analogous compatibility conditions:
\be
\mathcal{B}(j_{\mu(s)}(C,C))\approx0\,.
\ee
where the stationary surface is defined by the linear equations of motions for the jet $C$.
In the standard examples of massless fields the compatibility condition is \emph{conservation}. In addition, non-trivial current interactions are naturally defined as equivalence classes:
\be
j_{\mu(s)}(C,C)\sim j_{\mu(s)}(C,C)+\mathcal{F}_{\mu(s)}(\mathcal{I}(C,C))\,.\label{equiv}
\ee
Here $\mathcal{I}(C,C)\in \bar{Q}_{\infty}$ parametrises the most general (admissible) redefinition of the field $\phi_{\mu(s)}$ quadratic in the jet $C$:
\be
\phi_{\mu(s)}\rightarrow \phi_{\mu(s)}+\mathcal{I}_{\mu(s)}(C,C)\,.
\ee
Notice that the eventual Bianchi identities of the kinetic operator ensure that $\mathcal{F}_{\mu(s)}(\mathcal{I}(\phi,\phi))$ is an off-shell compatible current. Such currents are usually referred to in the physics literature as \emph{improvements} and parametrise the freedom in choosing different field frames of the theory.

To summarise, current interactions are defined by the following ``cohomological'' problem:
\be
\mathcal{B}(j_{\mu(s)}(C,C))\approx 0\,,\qquad j_{\mu(s)}\sim j_{\mu(s)}+\mathcal{F}_{\mu(s)}(\mathcal{I}(C,C))\,,
\ee
where with a small abuse of notation we include also the $s=0$ case where the current degenerates to a scalar bilinear.
In standard situations one restricts the above problem to local functionals: $j_{\mu(s)}(C,C)\,,\ \mathcal{I}_{\mu(s)}(C,C)\in \bar{Q}_{k\leq\infty}$. A standard example of non-trivial current interaction is then given by the standard electromagnetic current $j_\mu=i\phi^* \overset{\leftrightarrow}{\partial}\phi$ or the stress tensor, which encodes the backreaction of the scalar on the gravitational sector. Generalisations of these currents are known also for HS and follow from the Metsaev classification in flat space \cite{Metsaev:2005ar} and generalisations thereof to AdS space \cite{Vasilev:2011xf,Joung:2011ww,Boulanger:2012dx,Joung:2013doa}.

\paragraph{From local to non-local:} Given the setting described here-above, the main problem to address in a theory involving couplings with infinitely many derivatives like string theory or Vasiliev's equations is to characterise the most general admissible field redefinitions $\mathcal{I}_{\mu(s)}(C,C)\in \tilde{Q}_{\infty}$, beyond the strict locality condition.

Redefinitions of this type are indeed unavoidable if one wants to bring string-field theory interactions to their canonical Metsaev frame. Enlarging the functional space can however have undesired consequences like for instance the loss of any non-trivial solution. It has been already observed some time ago, for the stress tensor\cite{Prokushkin:1999xq}, and proved more recently in general\cite{Kessel:2015kna,Boulanger:2015ova,Skvortsov:2015lja}, that allowing arbitrary pseudo-local redefinitions in maximally symmetric backgrounds erases the difference between non-trivial local currents and improvements.
The prototypical example of such a scenario is a system of equations of the type:
\begin{subequations}
\begin{align}
(\Box-m^2)\Psi(x)&=\Phi^2(x)\,,\label{boxphi}\\
(\Box-\widetilde{m}^2)\Phi(x)&=0\,.
\end{align}
\end{subequations}
\newpage
\noindent In this case:
\begin{itemize}
\item Any local redefinition of $\Psi$ quadratic in $\Phi$ will contribute with interaction terms involving at least two derivatives. Hence the interaction term $\Phi^2$ is a non-trivial interaction when restricting the functional space to $\bar{Q}_{k<\infty}$.
\item Any term of the type $C_{\mu(s)}C^{\mu(s)}$ which can be inserted on the right-hand side of \eqref{boxphi} admits a completion to a local improvement of the type 
\be
C_{\mu(s)}C^{\mu(s)}-C_s \Phi^2\,,
\ee
where the coefficient depends on the mass $m$ of $\Psi$.
\item Allowing $\mathcal{I}\in{Q}_{\infty}$ one can find a trivial representative also for the $\Phi^2$ current:
\be
\mathcal{I}=\frac{1}{\Box-m^2}\,\Phi^2=-\frac1{m^2}\sum_{i=0}^\infty\left(\frac{\Box}{m^2}\right)^i\Phi^2\in Q^{\infty}\,.
\ee
\end{itemize}
What we have described above in a simple example is a generic feature for current interactions whenever the theory involves some mass parameter. In particular, one can show in various relevant cases on both AdS and flat space\cite{Kessel:2015kna,Boulanger:2015ova,Skvortsov:2015lja} that:
\begin{align}
\mathcal{B}(j_{\mu(s)}(C,C))\approx 0\,,\quad\Longrightarrow \quad j_{\mu(s)}=\mathcal{F}_{\mu(s)}(\mathcal{I}(C,C))\,,
\end{align}
with
\begin{align}
j_{\mu(s)}(C,C)\in \bar{Q}_\infty\,,\qquad \mathcal{I}(C,C)\in \bar{Q}_\infty\,,
\end{align}
The above statement also extends to massive fields on flat space\footnote{For massless fields in flat space the above does not apply since $\frac1{\Box}\notin Q_\infty$}.

A consequence of the above statement is that in a large class of theories, including string theory and Vasiliev's equations, any solution to the conservation condition:
\be
\mathcal{B}(j_{\mu(s)}(C,C))\approx 0\,,
\ee 
can be written as an infinite sum of \emph{local} conserved currents. By definition:
\begin{multline}
\mathcal{I}_{\mu(s)}(C,C)\in \bar{Q}_{\infty}\\\Longrightarrow\ \mathcal{I}_{\mu(s)}(C,C)=\sum_{n,m}(f_{JK})_{\mu(s)}{}^{\nu(n),\rho(m)}\,C^J_{\nu(m)}(x)C^{K}_{\rho(n)}(x)\,.
\end{multline}
and one easily concludes that
\begin{multline}
j_{\mu(s)}=\mathcal{F}_{\mu(s)}\left(\sum_{n,m}(f_{JK})_{\mu(s)}{}^{\nu(n),\rho(m)}\,C^J_{\nu(m)}(x)C^{K}_{\rho(n)}(x)\right)\\
=\sum_{n,m}\underbrace{\mathcal{F}_{\mu(s)}\left((f_{JK})_{\mu(s)}{}^{\nu(n),\rho(m)}\,C^J_{\nu(m)}(x)C^{K}_{\rho(n)}(x)\right)}_{\quad\quad\in \ \bar{Q}_{m+n+2}}\,.\label{nononlocal}
\end{multline}
The above statement is a key step in the analysis of conserved currents in $\bar{Q}_\infty$. It shows that there is \emph{no} intrinsically pseudo-local conserved current which cannot be written as sum of \emph{local} conserved currents. Hence, this result ensures the existence of a basis of local currents for the subspace $\mathcal{C}_{\infty}=\{j\in \bar{Q}_{\infty}\,\big|\,\mathcal{B}(j)\approx0\}$. Furthermore, this property ensures that starting from $\mathcal{C}_{k}=\{j\in \bar{Q}_{k}\,\big|\,\mathcal{B}(j)\approx0\}$ there exist continuous maps:
\be
\pi_{\infty,k}: \mathcal{C}_{\infty}\rightarrow \mathcal{C}_{k}\,,\quad \text{s.t.}\quad\forall j\in \mathcal{C}_{\infty}\,,\quad \pi_k(j)\in \mathcal{C}_{k}\,.
\ee
The maps $\pi_{\infty,k}$ can be chosen such that:
\begin{equation}\label{piC}
\pi_{\infty,k}\left(C_{\mu(m)}C_{\nu(n)}\right)=\left\{\begin{matrix}
C_{\mu(m)}C_{\nu(n)}&\quad\text{if}\,\quad n+m\leq k\\
0\,\,&\quad\text{if}\,\quad n+m>k
\end{matrix}\right.\,,
\end{equation}
which is the natural definition induced by the corresponding one for the scalar jet vector $C$, upon taking its tensor product with itself.

\paragraph{A Functional Class Proposal:}
So far we have gathered the needed ingredients in order to define our functional class proposal. Note that while no non-trivial current exist in $\bar{Q}_\infty$ they do exist in $\bar{Q}_k$. In the following we will call canonical currents the representatives for the non-trivial currents in $\bar{Q}_k$. Metsaev classification and its extensions to AdS imply that there always exists a representative with at most $s+s_1+s_2$ derivatives where $s$ is the spin of the current and $s_1$ and $s_2$ are the spins of the fields involved. We will refer to the latter representatives as $\mathbf{j}$. Furthermore for each element $j\in \mathcal{C}_k$ the following decomposition is unique and well defined:
\be
\forall j\in \mathcal{C}_k:\quad j=b_j \mathbf{j}+\mathcal{F}_{\mu(s)}(\mathcal{I}(C,C))\,,\quad \mathcal{I}(C,C)\in \bar{Q}_{k-2}\,,
\ee
and allows to extract the projection of each local current on the non-trivial cohomology representatives. Therefore, given $j\in \mathcal{C}_\infty$ one can consider the following sequence of decompositions:
\be
\pi_{\infty,k}(j)=b^{(j)}_k \mathbf{j}+\mathcal{F}(\mathcal{I}(C,C))\,,
\ee
which is uniquely defined. Our construction then allows to define the above decomposition directly in $\mathcal{C}_{\infty}$ arriving at:
\be
j=\lim_{k\to\infty}\pi_{\infty,k}(j)=\left(\lim_{k\to\infty}b_k^{(j)}\right)\mathbf{j}+\mathcal{F}(\lim_{k\to\infty}\mathcal{I}_k(C,C))\,.
\ee
The inverse limit provides in this way a prescription to extract the coefficient of the local cohomology from any formal element $j\in \mathcal{C}_\infty$. It also provides with a criterion to study if the inverse limit of a formal element $j\in \mathcal{C}_{\infty}$ exists. We conclude this section by defining the space of \emph{trivial} pseudo-local currents (pseudo-local improvements) as those elements $j\in \mathcal{C}_\infty$ such that:
\be
\lim_{k\to\infty}b_k^{(j)}=0\,.\label{imprdef}
\ee
For each current $j\in \mathcal{C}_{\infty}$ with the above property we thus identify a subset of $\bar{Q}_\infty$ which is the functional space of allowed pseudo-local redefinitions.

\section{Examples}\label{sec:Examples}

This section is devoted to few tests and examples of the ideas described in this article. We first describe the string theory example decomposing the string vertex according to its canonical and improvement pieces and then move to AdS space. We also consider the implications of the functional class proposal introduced in the previous section on the simplest example of $\phi^4$ coupling in AdS which has been subject of recent studies\cite{Bekaert:2014cea,Bekaert:2015tva}.

\subsection{String Theory}

The string theory case is one of the most instructive. String theory indeed naturally produces pseudo-local couplings involving infinitely many derivatives. This feature can be appreciated by computing the off-shell open string effective-action in the form:
\be
S=\int \left[\tfrac12\left\langle \Phi|L_0|\Phi\right\rangle+\ldots\right]+\langle 0|\,\mathcal{V}\,|\Phi\rangle\otimes|\Phi\rangle\otimes |\Phi\rangle+\mathcal{O}(\Phi^4)\,,
\ee
with $\mathcal{V}$ the string vertex which can be computed at tree-level by standard techniques\cite{Sciuto:1969vz,DiVecchia:1986mb,Neveu:1986ai,Gross:1986ia,DiVecchia:1988hq,Moeller:2005ez,Taronna:2010qq,Sagnotti:2010at}. Above the $\ldots$ are the required terms which allow to recover the Virasoro lowest weight condition upon varying the quadratic action and which we will systematically omit in the following. The equations of motion following from the above action would then read:
\be
\left(\,L_0|\Phi\rangle+\ldots\right)+\mathcal{V}\,|\Phi\rangle\otimes |\Phi\rangle+\mathcal{O}(\Phi^3)=0\,.
\ee
The vertex $\mathcal{V}$ can be easily computed at cubic order and expressed in terms of Neumann coefficients by choosing a simple parametrisation for the coordinates around the vertex operators insertions: $\xi_i(y)=y-y_i$. In terms of the Neumann coefficients one arrives in full generality to the following well-known Gaussian type of expression:
\be
\mathcal{V}=|y_{12}y_{13}y_{23}|\langle 0|\exp\left[-\frac12\sum_{{\substack{1\leq i<j\leq3\\0\leq n,m\leq\infty}}} \alpha_n^{\mu(i)}N^{ij}_{nm}(y_{ij})\alpha_{m\mu}^{(j)}\right]\,.
\ee
Upon computing explicitly the Neumann coefficients for the parametrisation $\xi_i(y)=y-y_i$, and restricting the attention for simplicity to the first Regge trajectory $|\Phi\rangle=\tfrac1{s!}\,\Phi_{\mu(s)}\,\alpha_{-1}^\mu\ldots\alpha_{-1}^\mu|0\rangle$, one finally arrives to:
\be
\mathcal{V}=|y_{12}y_{13}y_{23}|\exp\Bigg\{\sum_{i\,\neq\, j}^3\Big[-\alpha^\prime\pl_{i}\cdot \pl_{j}\ln|y_{ij}|-i\sqrt{2\alpha^\prime}\,\frac{\alpha^{(j)}_1\cdot \pl_{i}}{y_{ij}}+\frac{\alpha^{(i)}_1\cdot \alpha^{(j)}_1}{y_{ij}^2}\Big]\Bigg\}\,.\label{FirstRegge}
\ee
For the first Regge trajectory the quadratic part also simplifies. Therefore, varying the above action one arrives to equations of motion of the type:
\begin{align}
\mathcal{F}_{\mu(s)}(\Phi)=(\Box-\tfrac{s-1}{\alpha^\prime})\Phi_{\mu(s)}+\ldots=J_{\mu(s)}(\Phi,\Phi)\,,
\end{align}
where we have dropped again unessential terms for our discussion proportional to traces and divergences of the fields. We have also considered a gauge fixing of the Stueckelberg gauge symmetries. The current $J_{\mu(s)}$ can be extracted from \eqref{FirstRegge} and reads:
\begin{align}\label{stringCurrent}
J(x,\alpha_{-1}|\Phi,\Phi)&=e^{-\alpha^\prime \Box\ln|\frac{y_{12}}{y_{31}y_{23}}|+i\sqrt{\frac{\alpha^\prime}{2}}\,\frac{y_{12}}{y_{23}y_{31}}\alpha_{-1}\cdot \pl_{12}}\\
&\hspace{-10pt}\times {}_{12}\langle0|e^{+\frac{y_{12}}{y_{23}y_{31}}\alpha_{-1}\cdot (\alpha^{(1)}_1+\alpha^{(2)}_1)+i\sqrt{2\alpha^\prime}\left(\alpha^{(1)}_{1}\cdot \pl_{2}-\alpha^{(2)}_{1}\cdot \pl_{1}\right)+\alpha^{(1)}_1\cdot \alpha^{(2)}_1}\nonumber\\
&\hspace{-10pt}\times \,|\tfrac{y_{23}y_{31}}{y_{12}}|\,\Phi(x_1|\alpha^{(1)}_{-1})\Phi(x_2|\alpha^{(2)}_{-1})|0\rangle_{12}\Big|_{x_1=x_2=x}\,.\nonumber
\end{align}
Notice that after varying the action and going on-shell we can use the equations of motion for the fields $\Phi_1$ and $\Phi_2$ which from now on will be considered on their linear mass-shell as in the discussion of Section \ref{sec:pseudo}. However, even after going on-shell at the EoMs level the current still remains pseudo-local. Only a pseudo-local redefinition would possibly remove the non-localities!

This makes manifest the pseudo-local nature of the string interactions. Importantly, one also recovers some off-shell ambiguities parametrised by the insertion positions\footnote{Notice that if we had not fixed the coordinate parametrisation around each insertion position the string vertex would have explicitly depended on the choice of parametrisation.} $y_i$.

In the following we are going to apply the functional class proposal discussed in this article to decompose the above currents and identify the corresponding improvement and canonical parts.

\paragraph{Decomposition in Canonical plus Improvements: } As we have explained in Section \ref{sec:pseudo} we can achieve a well-defined decomposition of any interaction by a limiting procedure. In the end this reduces the problem to study a basis of infinitely many local elements. On the other hand, improvement terms have the generic form:
\be
\mathcal{F}_{\mu(s)}(\mathcal{I}(\Phi,\Phi))=(\Box-\tfrac{s-1}{\alpha^\prime})\mathcal{I}(\Phi,\Phi)+\ldots\,,
\ee
and by definition satisfy the Bianchi identities of the theory. The limiting procedure described in Section \ref{sec:pseudo} is here further simplified by the fact that derivatives commute each other. The improvement decomposition for a spin-$s$ current of the type:
\be
(\Box-m^2)\Phi_{\mu(s)}=\Box^n\Phi\pl_{\mu(s)}\Phi\,,
\ee
is then given by factorising the equations of motion as:
\begin{multline}
(\Box-m^2)\Phi_{\mu(s)}=(m^2)^{n}\Phi\pl_{\mu(s)}\Phi\\+(\Box-m^2)\sum_{i=1}^{n}\binom{n}{i}(\Box-m^2)^{i-1}(m^2)^{n-i}\,\Phi\pl_{\mu(s)}\Phi\,.
\end{multline}
Here we have explicitly disentangled the improvement piece.
One can then first expand the string current, perform the above decomposition and resum it. The final result is remarkably simple and reads:
\begin{align}
J(x,\alpha_{-1}|\Phi,\Phi)&=e^{-\alpha^\prime (\Box-\tfrac{\hat{N}-1}{\alpha^\prime})\ln|\frac{y_{12}}{y_{31}y_{23}}|}e^{i\sqrt{\frac{\alpha^\prime}{2}}\,\alpha_{-1}\cdot \pl_{12}}\\
&\hspace{-10pt}\times {}_{12}\langle0|e^{\alpha_{-1}\cdot (\alpha^{(1)}_1+\alpha^{(2)}_1)+i\sqrt{2\alpha^\prime}\left(\alpha^{(1)}_{1}\cdot \pl_{2}-\alpha^{(2)}_{1}\cdot \pl_{1}\right)+\alpha^{(1)}_1\cdot \alpha^{(2)}_1}\nonumber\\
&\hspace{-10pt}\times \,\Phi(x_1|\alpha^{(1)}_{-1})\Phi(x_2|\alpha^{(2)}_{-1})|0\rangle_{12}\Big|_{x_1=x_2=x}\,.\nonumber
\end{align}
The conclusion is then that the canonical current piece is given by:
\begin{multline}
J^{\text{can.}}(x,\alpha_{-1}|\Phi,\Phi)=e^{i\sqrt{\frac{\alpha^\prime}{2}}\,\alpha_{-1}\cdot \pl_{12}+i\sqrt{2\alpha^\prime}\left(\alpha^{(1)}_{1}\cdot \pl_{2}-\alpha^{(2)}_{1}\cdot \pl_{1}\right)+\alpha^{(1)}_1\cdot \alpha^{(2)}_1}\\
\times \,\Phi(x_1|\alpha^{(1)}_{-1})\Phi(x_2|\alpha^{(2)}_{-1})|0\rangle_{12}\Big|_{\substack{x_1=x_2=x\\ {\alpha_{-1}^{(i)}=\alpha_{-1}}}}\,,\nonumber
\end{multline}
and is local given any triple of spins matching Metsaev's classification. On the other hand the improvement current is pseudo-local even for fixed spin and is dressed by an exponential of $\Box$:
\begin{multline}
J^{\text{Impr.}}(x,\alpha_{-1}|\Phi,\Phi)=\left(e^{-\alpha^\prime (\Box-\tfrac{\hat{N}-1}{\alpha^\prime})\ln|\frac{y_{12}}{y_{31}y_{23}}|}-1\right)\\e^{i\sqrt{\frac{\alpha^\prime}{2}}\,\alpha_{-1}\cdot \pl_{12}+i\sqrt{2\alpha^\prime}\left(\alpha^{(1)}_{1}\cdot \pl_{2}-\alpha^{(2)}_{1}\cdot \pl_{1}\right)+\alpha^{(1)}_1\cdot \alpha^{(2)}_1}
\,\Phi(x_1|\alpha^{(1)}_{-1})\Phi(x_2|\alpha^{(2)}_{-1})|0\rangle_{12}\Big|_{\substack{x_1=x_2=x\\ {\alpha_{-1}^{(i)}=\alpha_{-1}}}}\,.\nonumber
\end{multline}
For instance, the spin-1 current built out of tachyons reads:
\be
J^\mu\sim \,e^{-\alpha^\prime\Box
\ln|\tfrac{y_{23}}{y_{31}y_{12}}|}
\Phi^\star\overset{\leftrightarrow}{\partial^\mu}\Phi
=\Phi^\star\overset{\leftrightarrow}{\partial^\mu}\Phi-
\alpha^\prime
\left(\ln|\tfrac{y_{23}}{y_{31}y_{12}}|\right)
\Box\,(\Phi^\star\overset{\leftrightarrow}{\partial^\mu}\Phi)
+\ldots\,.
\ee
We conclude this section by stressing that all the dependence on the insertion points, as well as any dependence on the parametrisation around each insertion, just appears in the improvement and defines the corresponding field frame on the effective field theory side. Different choices of parametrisation or insertion positions then just correspond to different infinite derivative tails for each vertex. The infinite derivative improvement tail appears from this perspective as a key ingredient to achieve a string interpretation of the underlying (effective) field theory. We conclude this section by stressing that the improvement term is an improvement according to the definition \eqref{imprdef}. Notice also that any higher-derivative term in eq.~\eqref{stringCurrent} contributes to the canonical component! In particular all of them are needed to achieve the cancellation of any dependence on the $y_i$ in the canonical current sector.

\subsection{Scalar Currents in $AdS_3$ and $AdS_4$}

The limiting procedure described in Section~\ref{sec:pseudo} is more subtle in curved backgrounds because of the non-commutative nature of covariant derivatives. The first step is to fix a basis for the jets of the fields. A convenient choice, which is natural in Vasiliev's theory is to consider symmetrised and traceless derivatives:
\be
C_{a(k)}(x)=i^kh_a^\mu\ldots h_a^\mu\nabla_{\mu\{k\}}\Phi(x)\,.
\ee
The above jet components are precisely the field variables entering the unfolded equations. Furthermore, due to convenient isomorphisms for the isometry algebras $so(2,2)\sim sp(2)\oplus sp(2)$ and $so(3,2)\sim sp(4)$, the above totally symmetric tensors can be encoded into:
\begin{align}
C_{\ga(2k)}(x)&\sim (ih^\mu_{\ga\ga}\nabla_{\mu})^k\Phi(x)\,,& C_{\ga(k)\gad(k)}(x)&\sim (ih^\mu_{\ga\gad}\nabla_\mu)^k\Phi(x)\,.
\end{align}
Here we have used the standard conventions expressing the vielbein in terms of Pauli matrices as
\begin{align}
h^{\ga\ga}_{\mu}&=\frac1{2z}\,(1,\sigma_3,\sigma_1)_\mu^{\ga\ga}\,,& h^{\ga\gad}_{\mu}&=\frac1{2z}(1,\sigma_3,\sigma_1,\sigma_2)_\mu^{\ga\gad}\,.
\end{align}
The tracelessness condition becomes in spinorial language a simple consequence of symmetrisation of the spinorial indices. These are indeed contracted with the antisymmetric $sp(2)$ invariant tensor\footnote{In the $sp(2)$ language we raise and lower indices as $A^\ga=\epsilon^{\ga\gb}A_\gb$ and $A_{\ga}=A^{\gb}\epsilon_{\gb\ga}$.} $\epsilon_{\ga\gb}$ ($\epsilon_{12}=1$).
The spinorial language is also convenient to express a basis for $\bar{Q}_\infty$ as:
\begin{align}
j_{\ga(2s)}&=C_{\ga(2s-2k)\nu(2l)}C^{\nu(2l)}{}_{\ga(2k)}\,,\\
j_{\ga(s)\gad(s)}&=C_{\ga(s-k)\nu(l)\gad(s-k)\dot{\nu}(l)}C_{\ga(k)}{}^{\nu(l)}{}_{\gad(k)}{}^{\dot{\nu}(l)}\,.
\end{align}
In the following, for ease of notation and simplicity, we will concentrate on a conformally coupled scalar because in this case the corresponding HS currents can be improved to traceless currents:
\begin{align}
d&=3\,,& (\Box+\tfrac34)\Phi(x)&=0\,,\\
d&=4\,,& (\Box+2)\Phi(x)&=0\,.
\end{align}
The above mass shell conditions are equivalent to the following recursion relations for the jet variables:
\begin{align}
\nabla C^{\ga(2k)}-ih^{\ga\ga}C^{\ga(2k-2)}+\tfrac{i}{2}\,h_{\gamma\gamma}C^{\gamma\gamma\ga(2k)}&=0\,,\\
\nabla C^{\ga(k)\gad(k)}-\tfrac12\, h^{\ga\gad}C^{\ga(k-1)\gad(k-1)}-h_{\gamma\dot{\gamma}}C^{\gamma\ga(k)\dot{\gamma}\gad(k)}&=0\,.
\end{align}
These allow to express covariant derivatives in terms of the other on-shell jet variables.
In the spinorial language the corresponding quadratic equations read:
\begin{align}
\mathcal{F}_{\ga(2s)}&=[\Box-s(s-2)]\phi_{\ga(2s)}+\ldots=j_{\ga(2s)}\,,\\
\mathcal{F}_{\ga(s)\gad(s)}&=[\Box-(s-2)(s+1)]\phi_{\ga(s)\gad(s)}+\ldots=j_{\ga(s)\gad(s)}\,.
\end{align}
The assumption of a conformally coupled scalar simplifies also the current conservation condition which usually involves both the current and its trace. In our case the conservation conditions for currents sourcing the Fronsdal tensor read as usual:
\begin{align}
\nabla^{\gb\gb}j_{\gb\gb\ga(2s-2)}&\approx0\,,\\
\nabla^{\gb\gbd}j_{\gb\ga(s-1)\gbd\gad(s-1)}&\approx0\,,
\end{align}
respectively in 3 and 4 dimensions. The conservation condition can be solved in general\cite{Prokushkin:1999xq,Gelfond:2006be,Gelfond:2010pm,Gelfond:2014pja,Kessel:2015kna,Boulanger:2015ova,Skvortsov:2015lja} and admits the following solution in the space $\bar{Q}_\infty$ of pseudo-local currents.
\paragraph{3d:} In 3d the solution to current conservation reads:
\be
j_{\ga(2s)}=\sum_{l=0}^\infty\alpha_{l}\,j^{(l)}_{\ga(2s)}\,,\label{3dcurrent}
\ee
expressed in terms of the following local basis of independently conserved currents:
{\small
\begin{align}
j^{(2l)}_{\ga(2s)}&=\frac{(-1)^{l}}{(2l)!(2s)!}\sum_{k=0}^{s}(-1)^k\binom{2s}{2k}\,C_{\ga(2s-2k)\nu(2l)}(x)C^{\nu(2l)}{}_{\ga(2k)}(x)\,,\\
j^{(2l+1)}_{\ga(2s)}&=i\frac{(-1)^{l}}{(2l+1)!(2s)!}\sum_{k=0}^{s-1}(-1)^{k+1}\binom{2s}{2k+1}\,C_{\ga(2s-2k-1)\nu(2l+1)}(x)C^{\nu(2l+1)}{}_{\ga(2k+1)}(x)\,.
\end{align}}
\paragraph{4d:} In 4d the solution to current conservation reads:
\be
j_{\ga(s)\gad(s)}=\sum_{l=0}^\infty\alpha_{l}\,j^{(l)}_{\ga(s)\gad(s)}\,,\label{4dcurrent}
\ee
expressed in terms of the following local basis of independently conserved currents:
{\small
\begin{align}
j^{(l)}_{\ga(s)\gad(s)}&=(-1)^l\,\frac{1}{(l!)^2(s!)^2}\sum_{k=0}^{s}(-1)^k\binom{s}{k}^2\,C_{\ga(s-k)\nu(l)\gad(s-k)\dot{\nu}(l)}(x)C_{\ga(k)}{}^{\nu(l)}{}_{\gad(k)}{}^{\dot{\nu}(l)}(x)\,.
\end{align}}

As anticipated, any pseudo-local conserved current $j\in Q_\infty$ can be written as (infinite) sum of local conserved currents. Therefore, the above result shows explicitly that for any current $j\in Q_\infty$ such that $\nabla\cdot j=0$ then $j\in \mathcal{C}_\infty$. The above is just a convenient choice of basis for each $\mathcal{C}_k$ which is also convenient to study the action of the $\pi_{\infty,k}$-maps \eqref{piC}.

\paragraph{Decomposition in canonical plus improvements: }

The decomposition described in Section~\ref{sec:pseudo} amounts in the AdS case to a decomposition in improvement and canonical part for each element of the basis introduced above. Given any element of $\mathcal{C}_\infty$ one can indeed first decompose it in this basis. One then finds the corresponding decomposition of the associated local truncations simply from the decomposition of the basis elements.

By using the explicit form of the Fronsdal operator in spinorial notation it is possible to find the most general form of a local traceless improvement\cite{Skvortsov:2015lja}. They read in terms of the basis introduced above:
\begin{align}
\mathcal{I}_s^{(l)}=j_s^{(l-1)}-C^{(s)}_l\,j_s^{(0)}\,.
\end{align}
The coefficients $C^{(s)}_l$ read in 3d:
\begin{multline}\label{Ccoeffs}
C^{(s)}_l=(-1)^l\frac{ s\, (2 s+l)!}{(2 s)! (l+1)!} \Big[2 \,(l+s) \, _2F_1(1,l+2 s+1;l+2;-1)-l-1\Big]\\+4^{-s} (l+s)\,,
\end{multline}
and in 4d:
\be
C^{(s)}_l=-\frac{(-1)^{l}}{2 (2 s-1) }\frac{ \,(l+s)!}{ s! (s-1)!\Gamma(1+l-s)}\,{}_3F_2\left(
\begin{matrix}
1-s\quad1-s\quad-2 s\\
2-2 s\quad1+l-s
\end{matrix};1\right)\,.
\ee
Their asymptotic behaviour for $l\rightarrow\infty$ is:
\begin{subequations}
\begin{align}
C^{(s)}_l&\sim l^{2s-1}\,,&d&=3\,,\\
C^{(s)}_l&\sim l^{2s}\,,&d&=4\,.
\end{align}
\end{subequations}
Notice that in the scalar case under consideration there is one local cohomology and one can choose a representative for it given precisely by $j^{(0)}_s$. The coefficient $C_l^{(s)}$ is precisely the cohomology coefficient of the decomposition of each $j^{(l)}_s$. The above spin-dependent behaviour shows a key difference among the flat and the AdS examples. In the latter the non-commutative nature of covariant derivatives shows a factorially growing behaviour for the contribution to the canonical current sector of the higher-derivative terms.

In this setting one can also compute the decomposition into canonical and improvement pieces of any element in $\mathcal{C}_\infty$ from:
\be
\pi_{\infty,s+k}(j_s)=\sum_{l=0}^{k}\alpha_l\,j_s^{(l)}=\underbrace{\left(\sum_{l=0}^k\alpha_l\,C_{l+1}^{(s)}\right)}_{b_k}j^{(0)}_s+\text{Impr.}\,,
\ee
and taking the limit:
\be
b=\lim_{k\to\infty}b_k=\sum_{l=0}^\infty\alpha_l\,C_{l+1}^{(s)}\,.\label{limtb}
\ee
The existence and the result of the above limit allows to define accordingly well-defined pseudo-local currents (when the limit exists) and Improvements (when the limit exists and is zero).

\paragraph{Witten diagram computation in 3d and holographic checks of the functional class proposal: } The simplest check of the functional class criterion proposed above is the holographic check\cite{Skvortsov:2015lja}. One can indeed compute the Witten diagrams associated to the higher-derivative terms forming a basis for $\mathcal{C}_\infty$. The 3d computation can be done by standard techniques\cite{Chang:2011mz,Chang:2011vka,Ammon:2011ua,Fitzpatrick:2014vua,Alkalaev:2015wia,Skvortsov:2015lja} and reads:
\begin{equation}\label{CFT}
\int_{AdS_3}\phi^{\ga(2s)}j^{(l)}_{\ga(2s)}=\pi\,C^{(s)}_{l+1}\,\left(-\frac1{4}\right)^{s}\frac{(s-1)!}{|x_{12}|}\left(\frac{x_{12}^+}{x_{13}^+x_{23}^+}\right)^s\,.
\end{equation}
where the $C^{(s)}_l$ coincide with the coefficients given in eq.~\eqref{Ccoeffs}. Therefore, the Witten diagram computation leads to the same functional class criterion proposed in this note. This result also shows that our criterion for the functional class in HS theory gives a prescription for a proper set of non-local redefinitions that \emph{do not} affect the computation of AdS/CFT observables (Witten diagrams).

Furthermore, from the AdS/CFT perspective a key aspect of the above computation that should be stressed is that the result of the Witten diagram for a pseudo-local interaction follows from \eqref{CFT} upon commuting the infinite sum over derivatives and the integral over AdS space. The locality criterion proposed in this paper turns out to be equivalent to the condition that the infinite sum over derivatives commutes with the integral over AdS-space.

\subsection{$\phi^4$ couplings: an argument using cubic locality}\label{quarticlocality}
Before concluding we would like to try and push our analysis to the simplest cubic coupling at the EoMs level: the one involving three scalar fields as sources to the scalar EoMs. This coupling has been recently subject of investigations\cite{Bekaert:2014cea,Bekaert:2015tva}. Let us stress that the analysis described here is preliminary and it is presented as an example of how similar issues as those described at cubic level are expected to manifest at higher orders.

In more details, in a theory of scalars there are infinitely many non-trivial cubic sources which can be written down. A simple way to perform the counting is to count inequivalent combinations of Mandelstam variables in flat space. One can then use the condition $s+t+u=\text{const.}$ to reduce the number of independent structures to:
\begin{align}
&s^n t^m\,,& n&\geq m\,,
\end{align}
where bose symmetry has been used to obtain the inequality. Back to coordinate space, the above basis can be conveniently recast into the following basis:
\begin{align}\label{scalarInt}
\mathcal{V}^{(s,l)}_4&=C^{\ga(s)\gad(s)} j_{\ga(s)\gad(s)}^{(l)}\,,& &s=2n,\quad n\geq l\geq0\,,\quad n,l\in \mathbb{N}\,.
\end{align}
We stress again that the above should be considered as a basis for the cubic scalar sources to the scalar field equations.

On the other hand, coupling the scalar to a tower of HS fields:
\begin{subequations}\label{cubic}
\begin{align}
\mathcal{F}_{\ga(s)\gad(s)}(\phi_s)&=j^{(0)}_{\ga(s)\gad(s)}\,,\\
(\Box+2)\Phi(x)&=\sum_{s=0}^\infty \phi_{\ga(s)\gad(s)}C^{\ga(s)\gad(s)}+\sum_{s,l} \alpha_{s,l}\,j_{\ga(s)\gad(s)}^{(l)}\,C^{\ga(s)\gad(s)}\,,
\end{align}
\end{subequations}
entangles the quartic self interaction of the scalar with the cubic interactions studied in the previous section\footnote{Above we have also introduced an auxiliary scalar $\phi_0$ with mass $m^2$ and corresponding interactions with the original scalar $\Phi$ on which $j$ depends upon. Although this additional scalar is not present in Vasiliev's equation $\phi_0$ can be decoupled in the limit $m^2\to \infty$. Introducing this auxiliary scalar is convenient at the perturbative level to argue about locality of quartic interactions.}. At this order any redefinition of the type studied at the cubic level would induce a cubic scalar interaction of the type in eq.~\eqref{scalarInt}. We then conclude that at the quartic order the interactions in \eqref{scalarInt} can be removed up to a suitable choice of improvement interactions\footnote{This is not strictly true in the presence of quadratic spin-s sources to the scalar equations. However, we believe this subtlety to not invalidate our argument about locality since locality of a quartic vertex should not be influenced by the addition of other cubic interactions. Furthermore, considering a formal expansion parameter $\lambda$ weighting the redefinition, the contribution to the quartic vertex generated by a quadratic spin-s source will be order $\lambda^2$ and hence subleading in the $\lambda\to 0$ limit.}. These would be generated by the redefinition removing the quartic contact terms and would still contribute to the quartic exchange amplitude\cite{Taronna:2011kt,Bekaert:2014cea}.

We can then reiterate the whole philosophy of this note applying it to the sum over $s$ and $l$ in \eqref{scalarInt}. We first truncate both $s$ and $l$ to finite values and remove the corresponding local interactions by a corresponding redefinition. At the end we take the limit $s,l\to \infty$.

Since we are interested in the analysis of redundant basis for the quartic interaction we drop the inequality $n\geq l\geq 0$. In this way, starting from \eqref{cubic} we generate a pseudo-local cubic theory whose equations of motion read:
\begin{subequations}
\begin{align}\label{cubic2}
\mathcal{F}_{\ga(s)\gad(s)}(\phi_s)&=\tilde{j}_{\ga(s)\gad(s)}(\Phi,\Phi)\,,\\
(\Box+2)\Phi(x)&=\sum_{s=0}^\infty \phi_{\ga(s)\gad(s)}C^{\ga(s)\gad(s)}\,,\\
&\hspace{-30pt}\tilde{j}_{\ga(s)\gad(s)}(\Phi,\Phi)\equiv j_{\ga(s)\gad(s)}^{(0)}(\Phi,\Phi)-\sum_{l=0}^\infty\alpha_{s,l}\,\mathcal{F}_{\ga(s)\gad(s)}\left(j^{(l)}_s(\Phi,\Phi)\right)\,,
\end{align}
\end{subequations}
It is then natural to require that the latter pseudo-local expression satisfies our locality proposal in the limit $s,l\to \infty$\footnote{A subtlety of this reasoning is that one may as well conclude that it is not possible to move the quartic non-local interaction to the cubic order. However, since this is possible for any local truncation thereof and for any local coupling in general, this would indicate a problem for the existence of the corresponding pseudo-local limit. In this note we argue that this behavior may be considered as a sign of non-locality of the quartic interaction.}.
Making use of the explicit form of the $\mathcal{F}_{\ga(s)\gad(s)}$ operator one can express the current $\tilde{j}_s$ in the basis \eqref{4dcurrent} obtaining for $d=4$ the following coefficients:
\be
\tilde{\alpha}_l^{(s)}=\delta_{l,0}-\frac{1}{4 (s-1)}\left[l^2 \alpha_{s,l-2}+\alpha_{s,l-1} \left(2 l s+2 (l+1)^2+s^2\right)+\alpha_{s,l} (l+s+2)^2\right]\,,
\ee
On can then argue that the coefficients in a cubic scalar source will be compatible with the definition of locality given in this note iff:
\be\label{quarticloc}
\sum_{l=0}^\infty\tilde{\alpha}^{(s)}_l C^{(s)}_l=1\,,
\ee
which is trivially satisfied for local interactions but quite non-trivial for pseudo-local ones. As an example of non-local interaction the following choice of coefficients:
\be
\alpha_{s,l}=\frac1{l^n}\,,\qquad n\geq1\,,
\ee
gives a non-local quartic interaction for any value of $n$. When the series in \eqref{quarticloc} is convergent but \eqref{quarticloc} is not satisfied, it is possible to ``improve'' the coefficients to make \eqref{quarticloc} satisfied by employing the exact representation for canonical currents\cite{Skvortsov:2015lja}:
\be
\alpha^{\Delta}_{s,l}=\Delta (-1)^l(2l+s+2)\left(\frac{l!}{(l+s+1)!}\right)^2\,,
\ee
which gives
\be
\tilde{\alpha}_{s,l}=(\Delta+1)\delta_{l,0}\,.
\ee
However, this is not possible when the series is divergent which happens when $n\leq2s+4$. We thus get the generic necessary but non sufficient condition:
\be
\alpha_{s,l}\prec\frac1{l^{2s+4}}\,.
\ee

The above results are argued to extend our locality condition at quartic order\cite{Bekaert:2015tva,Skvortsov:2015lja} and show new subtleties which may arise in discussing quartic locality. Remarkably, it seems a generic feature of pseudo-local quartic interaction to not be compatible with \eqref{quarticloc} even when the series in \eqref{quarticloc} converges\footnote{Similar features have been also appreciated in flat space\cite{Taronna:2011kt,Taronna:2012gb}.}. Notice that the basis \eqref{4dcurrent} includes factorial damping coefficients which do not suffice to ensure locality of any (convergent) infinite combination thereof. In view of these results it would be desirable to have a better understanding of the locality properties of quartic vertices appearing in HS theory.

\paragraph{Sum over spins vs. sum over derivatives:} An interesting interplay between sum over derivatives and sum over spins can be appreciated by expanding eq.~\eqref{scalarInt} as:

{\small
\begin{align}
\mathcal{V}_{s,l}&=\sum_{k+q=s}\frac{(-1)^{l+q+k}}{(l!)^2(k)!^2(q!)^2}\,C^{\ga(k)\ga(q)\gad(k)\gad(q)}C_{\ga(k)\nu(l)\gad(k)\dot{\nu}(l)}C_{\ga(q)}{}^{\nu(l)}{}_{\gad(q)}{}^{\dot{\nu}(l)}\,.
\end{align}}

\noindent
One can then rewrite the same basis elements as:

{\small
\begin{multline}
\mathcal{V}_{s,l}=\sum_{k+q=s}\frac{(-1)^{k}}{(k)!^2}C^{\ga(q)\nu(l)\gad(q)\dot{\nu}(l)}\left[(-1)^{q+l}\frac{1}{q!^2l!^2}\,C^{\ga(k)}{}_{\ga(q)}{}^{\gad(k)}{}_{\gad(q)}C_{\ga(k)\nu(l)\gad(k)\dot{\nu}(l)}\right]\\
=\sum_{k+q=s}C^{\ga(q+l)\gad(q+l)}I^{(k)}_{\ga(q+l)\gad(q+l)}\,.
\end{multline}}

\noindent
The latter cubic source can then be redefined by a redefinition of the spin-$(q+l)$ field analogously as in the discussion here-above. The only difference is that the corresponding redefinition will in this case generate also tracefull improvements at the cubic order. Dropping those\cite{Skvortsov:2015lja}, one can still argue some condition for locality projecting everything on the canonical current sector. This time one can work at fixed $q+l$ obtaining a necessary condition for the $s\to \infty$ limit of the coefficients $\alpha_{s,l}$. In general, given the following redundant manifestly cyclic basis:
\be
\mathcal{V}=\sum_{k,l,q}\,\frac{\alpha_{k,l,q}(-1)^{l+q+k}}{(l!)^2(k)!^2(q!)^2}\,C^{\ga(k)\ga(q)\gad(k)\gad(q)}C_{\ga(k)\nu(l)\gad(k)\dot{\nu}(l)}C_{\ga(q)}{}^{\nu(l)}{}_{\gad(q)}{}^{\dot{\nu}(l)}\,,
\ee
we obtain that the corresponding infinite set of coefficients can be improved to local coefficients as discussed above whenever the following convergence behaviour is verified:
\begin{align}
\alpha_{k,l,q}&\prec \frac1{k^{2(l+q)+4}}\,, \qquad k\to\infty\,, l+q=\text{const.}\,,\\
\alpha_{k,l,q}&\prec \frac1{l^{2(q+k)+4}}\,, \qquad l\to\infty\,, q+k=\text{const.}\,,\\
\alpha_{k,l,q}&\prec \frac1{q^{2(k+l)+4}}\,, \qquad q\to\infty\,, k+l=\text{const.}\,.
\end{align}
Otherwise the corresponding series will be divergent and there will be no way to improve the coefficients by employing the exact representative for canonical currents.
The above basis gives a cyclic form of the convergence condition which shows some interplay between sum over derivatives and the sum over spin. In more physical terms, at the amplitude level, one can argue that what in the $s$-channel might appear as a non-locality in derivatives will appear in another channel as some divergent sum over exchanges (c.f. Section \ref{sec:CFT}). On the other hand, something that appears local for that concerns the sum over $l$ above may still be not admissible with respect to the sum over $k$ or $q$.

\section{Locality and AdS/CFT}\label{sec:CFT}

Bulk locality at scales below the AdS length is a key concept which plays an important role within the holographic framework of the AdS/CFT correspondence\cite{Polchinski:1999ry,Giddings:1999jq,Gary:2009ae,Heemskerk:2009pn,ElShowk:2011ag,Fitzpatrick:2012cg,Fitzpatrick:2012yx,Maldacena:2015iua}. In simple terms one deals with three length-scales: the AdS-length $L$, the string-length $l_s$ and the Planck-length $l_{p}$. Standard AdS/CFT identifications can be then summarised as:
\begin{align}
\left(\frac{L}{l_p}\right)^4&=N\,,& \left(\frac{L}{l_s}\right)^4&=\lambda\,,
\end{align}
with $\lambda=N g_{YM}^2$ the {'}t Hooft coupling and the identification of the string coupling constant as $g_s=g_{YM}^2$. The standard AdS/CFT lore then requires a large-$N$ limit in order to recover a classical bulk description. Locality at sub-AdS distances is instead governed by the {'}t Hooft coupling constant. In particular, in the limit $\lambda\to \infty$ one would recover a local bulk-description. This limit requires strong coupling on the CFT side. In the HS case one is however interested into the opposite limit $\lambda\to0$ with $g_{YM}\to 0$ so that $g_s\to0$ and both the boundary and the bulk side are weakly coupled. In this limit the string-length would satisfy $l_s\gg L$ and one naturally recovers the tensionless limit of string theory. HS theories are then conjectured to be an effective description of the bulk side of the correspondence in this regime\cite{Sezgin:2002rt,Klebanov:2002ja}. However, due to the inequality $l_s\gg L$ one cannot expect to recover the same local bulk description which was available in the $\lambda\to \infty$ limit. We thus expect a generalisation of standard concepts of locality to be needed in order to consistently describe the bulk side in this regime. In this section we summarise some discussions carried at the conformal field theory level and their possible implications for bulk locality in the cases relevant for HS theories. 

\subsection{Mellin Amplitudes and Higher-Spin Theories}
The Mellin representation\cite{Mack:2009mi} for conformal field theory correlator has shown to be very useful to study bulk-locality in the context of AdS/CFT. Indeed poles in the Mellin amplitude are associated to bulk exchanges in a remarkable analogy with flat space S-matrix\cite{Penedones:2010ue,Fitzpatrick:2011ia,Paulos:2011ie}. In this language locality is translated into the analytic structure of Mellin amplitudes and in the corresponding singularities in analogy with S-matrix theory\cite{Fitzpatrick:2011hu,Fitzpatrick:2011dm}.

\paragraph{The Mellin amplitude of the $O(N)$ vector model:} In the free theory case CFT correlators can be easily computed by Wick-product. Computing the corresponding Mellin amplitude then amounts to invert the Mellin transform\cite{Mack:2009mi}:
\begin{subequations}
\begin{align}\label{mellin}
\left\langle\mathcal{O}_{\Delta}(x_1)\mathcal{O}_{\Delta}(x_2)\mathcal{O}_{\Delta}(x_3)\mathcal{O}_{\Delta}(x_4)\right\rangle&=\frac{\mathcal{N}}{(2\pi i)^2}\int d\delta_{ij}\,\tilde{M}(\delta_{ij})\,\prod_{i<j}^4x_{ij}^{-2\delta_{ij}}\,,\\
\tilde{M}(\delta_{ij})&=M(\delta_{ij})\prod_{i<j}^4\Gamma(\delta_{ij})\,,
\end{align}
\end{subequations}
where $\tilde{M}(\delta_{ij})$ is the Mellin transform of the CFT correlator, $M(\delta_{ij})$ is what is usually referred to as Mellin amplitude and $\mathcal{N}$ is some normalisation factor.
Above we introduced the Mellin variables $\delta_{ij}$ playing a similar role to Mandelstam variables in flat space. These variables satisfy:
\begin{align}
\delta_{ii}&=0\,,& \sum_{j}\delta_{ij}=\Delta_j\,,
\end{align}
with $\Delta_j$ the operator dimensions. The suggestive nature of the above conditions can become more apparent by introducing auxiliary ``momenta'' $k_i$ satisfying:
\begin{align}
k_i^2&=-\Delta\,,& \sum_i k_i&=0\,.
\end{align}
In the 4pt case and for equal conformal dimensions one can then define analogs of the Mandelstam variables as:
\begin{subequations}
\begin{align}
s&=-(k_1+k_2)^2=2\Delta-2\delta_{12}\,,\\
t&=-(k_1+k_3)^2=2\Delta-2\delta_{13}\,,\\
u&=-(k_1+k_4)^2=2\Delta-2\delta_{14}\,,
\end{align}
\end{subequations}
satisfying the standard relation $s+t+u=4\Delta$. In the following we apply the above formalism to the 4pt correlator of four scalar operators $\mathcal{O}=\tfrac1{\sqrt{2N}}\phi^a\phi_a$ of dimension $\Delta=d-2$ which is given by:
\begin{multline}
\left\langle\mathcal{O}_{\Delta}(x_1)\mathcal{O}_{\Delta}(x_2)\mathcal{O}_{\Delta}(x_3)\mathcal{O}_{\Delta}(x_4)\right\rangle=\frac{1}{(x_{12}^2x_{34}^2)^{\Delta}}+\frac{1}{(x_{13}^2x_{24}^2)^{\Delta}}+\frac{1}{(x_{14}^2x_{23}^2)^{\Delta}}\nonumber
\\
+\frac{4}{N}\left(\frac{1}{(x_{12}^2x_{34}^2)^{\frac{\Delta}{2}}(x_{13}^2x_{24}^2)^{\frac{\Delta}{2}}}+\frac{1}{(x_{13}^2x_{24}^2)^{\frac{\Delta}{2}}(x_{14}^2x_{23}^2)^{\frac{\Delta}{2}}}+\frac{1}{(x_{14}^2x_{23}^2)^{\frac{\Delta}{2}}(x_{12}^2x_{34}^2)^{\frac{\Delta}{2}}}\right)\,.
\end{multline}
Remarkably, although the Mellin transform of a function:
\be
f_M(\delta)\equiv\int_0^\infty dx\,x^{\delta-1}f(x)\,,
\ee
is ill defined when $f(x)=x^{\Delta}$, one can easily guess the corresponding result as a distributions by a simple inspection of \eqref{mellin}. It is indeed easy to show that the following distribution:
\begin{multline}\label{MellinOn}
\tilde{M}(s,t,u)\sim \delta\left(\tfrac{s}{2}\right)\delta\left(\tfrac{t}{2}-\Delta\right)\delta\left(\tfrac{u}{2}-\Delta\right)\\+\frac{4}{N}\delta\left(\tfrac{s}{2}-\tfrac{\Delta}{2}\right)\delta\left(\tfrac{t}{2}-\tfrac{\Delta}{2}\right)\delta\left(\tfrac{u}{2}-\Delta\right)+\text{cycl.}\,,
\end{multline}
solves \eqref{mellin} in the case of the free boson four-scalar correlator. Above we have introduced a third delta function and integration by using $s+u+t=4\Delta$ to arrive at a manifestly cyclic expression. One can recover a more standard expression depending upon two Mellin variables simply integrating on one among $s,t,u$.

Coming back to eq.~\eqref{mellin}, let us stress that the Mellin amplitude is usually defined as the Mellin transform of the CFT correlator up to some $\Gamma$ function factors. In the 4pt case this factor reads $M(\delta_{ij})=\tilde{M}(\delta_{ij})\left[\Gamma(\Delta-\tfrac{s}{2})\Gamma(\Delta-\tfrac{t}{2})\Gamma(\Delta-\tfrac{u}{2})\right]^{-2}$. These account for double-trace operators poles. Due to the singular structure \eqref{MellinOn} it is not clear to us how to treat the $\Gamma$-function factors whose arguments coincide with those of the $\delta$-functions.

To summarise, this discussion shows that the Mellin transform of the free $O(N)$ model is naturally a distribution and does not display any pole in the Mellin-variables. Notice also that the above $\delta$-functions are concentrated on points of the phase space which would not be allowed in the kinematical region for the generalised momenta $k_i$. This means that effectively the Mellin amplitude above \emph{vanishes} for ``physical'' parameters, possibly as a consequence of the HS symmetry.
Let us also stress that the above singular nature for the Mellin transform of the CFT correlator is a generic feature of CFTs with power-like dependence on the cross-ratios. This would remain true  for instance for the critical $O(N)$ model at large $N$, conjectured\cite{Klebanov:2002ja} to be dual to Vasiliev's equations with scalar boundary conditions $\Delta=2$. On the other hand these features do not appear in strongly coupled CFTs.

\paragraph{Sum over exchanges and locality:} The above result for the Mellin amplitude of the $O(N)$ model is expected, because the boundary theory is free, but it is also a bit puzzling in view of the HS dualities. From a bulk perspective one can indeed decompose the scattering process into an infinite number of HS exchanges which individually contribute with poles to the Mellin amplitude. This follows from the existence of non-trivial cubic couplings in the theory.
The contribution of a spin-s exchange would indeed have the following generic structure:
\be
M_n=\sum_m\frac{P^{(n)}_m(t)}{s-(\Delta_n+2m)}\,.
\ee
The sum over $m$ parameterises the sum over descendants of the single trace operator exchanged and the numerators are some polynomial in the Mellin variables which closely resemble the analogous polynomials in the Mandelstam variables appearing at the S-matrix level.

In a theory including infinitely many HS particles the full exchange amplitude would then involve an infinite sum over primaries of arbitrary spin:
\be
M=\sum_n g_n M_n=\sum_{n,m}\frac{g_n\,P^{(n)}_m(t)}{s-(\Delta_n+2m)}\,.
\ee
The full Mellin amplitude also involves contact terms\cite{Bekaert:2015tva} and the sum of those and the exchange\footnote{For exchanges involving external HS operators one can argue\cite{Taronna:2011kt,Taronna:2012gb} the current exchange contribution not to be gauge invariant. In this case a non-vanishing contact term would be required to get a gauge invariant amplitude.} should then be equal to the distribution in \eqref{MellinOn}. One possibility is to make an analogy with the case of conformal HS fields in flat space\cite{Joung:2015eny}.
The corresponding exchanges are given in this case by Gegenbauer polynomial and the sum over spins in the conformal HS case is indeed a distribution, which we quote\cite{Joung:2015eny} here:
\be
\sum_{s=0}^\infty(s+\tfrac{d-3}{2})\,C_s^{(\tfrac{d-3}{2})}\left(z\right)\sim\frac{(-1)^{d-4}}{(d-4)!}\delta^{[d-4]}\left(z\right)\,.
\ee
It is proportional to the $(d-4)$-th derivative of the $\delta$-function.

Although the analysis carried out here is incomplete, one may draw similar conclusions for the bulk contact term as for the flat space case\cite{Taronna:2011kt,Dempster:2012vw,Taronna:2012gb}. A pseudo-local behaviour may be expected in the regime under consideration where $\tfrac{L}{l_s}\ll1$. In view of the above results and of those of Section \ref{quarticlocality}, a better understanding of pseudo-locality in the bulk side at the quartic-order is a key step to improve our understanding of the HS-CFT dualities.

\section{Conclusions}\label{sec:Conclusions}

In this article we have studied locality in a field theory context with the aim of extending it to a quantum-gravity framework. In particular, we have formulated a functional class proposal to understand pseudo-local field redefinitions. Various examples have been considered and the functional class proposal has been tested in the string theory and in the AdS/CFT context. While the results of this article are encouraging and an extension of the locality proposal to the quartic order has been presented, more work is needed to achieve a better understanding of locality in the higher-spin framework both from a bulk and a boundary perspective. Hopefully, these investigation will shed some more light also on analogous questions in the context of Holography.

\section*{Acknowledgments}
\label{sec:Aknowledgements}

We would like to thank Glenn Barnich, Xavier Bekaert, Nicolas Boulanger, Slava Didenko, Maxim Grigoriev, Euihun Joung, Rakibur Rahman, Charlotte Sleight, Zhenya Skvortsov, Arkady Tseytlin and Mikhail Vasiliev for useful comments and discussions. We are also grateful to the organisers of the workshop ``Higher Spin Gauge Theorie'' at the Institute of Advanced Studies from Nanyang Technological University in Singapore for the opportunity to present this work. The research of M. Taronna is partially supported by the Fund for Scientific Research-FNRS Belgium, grant FC 6369 and by the Russian Science Foundation grant 14-42-00047 in association with Lebedev Physical Institute.

\bibliographystyle{ws-rv-van}
\bibliography{ws-rv-sample}

\begin{thebibliography}{69}
\providecommand{\natexlab}[1]{#1}
\providecommand{\url}[1]{\texttt{#1}}
\expandafter\ifx\csname urlstyle\endcsname\relax
  \providecommand{\doi}[1]{doi: #1}\else
  \providecommand{\doi}{doi: \begingroup \urlstyle{rm}\Url}\fi

\bibitem{1966asm..book.....E}
R.~J. {Eden}, P.~V. {Landshoff}, and D.~I. {Olive}, \emph{{The analytic
  S-matrix}}. 1966.

\bibitem{Heemskerk:2009pn}
I.~Heemskerk, J.~Penedones, J.~Polchinski, and J.~Sully, {Holography from
  Conformal Field Theory}, \emph{JHEP}. {\bf 10}, \penalty0 079,  (2009).
\newblock \doi{10.1088/1126-6708/2009/10/079}.

\bibitem{ElShowk:2011ag}
S.~El-Showk and K.~Papadodimas, {Emergent Spacetime and Holographic CFTs},
  \emph{JHEP}. {\bf 10}, \penalty0 106,  (2012).
\newblock \doi{10.1007/JHEP10(2012)106}.

\bibitem{Fitzpatrick:2012cg}
A.~L. Fitzpatrick and J.~Kaplan, {AdS Field Theory from Conformal Field
  Theory}, \emph{JHEP}. {\bf 02}, \penalty0 054,  (2013).
\newblock \doi{10.1007/JHEP02(2013)054}.

\bibitem{Maldacena:2015iua}
J.~Maldacena, D.~Simmons-Duffin, and A.~Zhiboedov, {Looking for a bulk point}.
  (2015).

\bibitem{Amati:1987wq}
D.~Amati, M.~Ciafaloni, and G.~Veneziano, {Superstring Collisions at Planckian
  Energies}, \emph{Phys. Lett.} {\bf B197}, \penalty0 81,  (1987).
\newblock \doi{10.1016/0370-2693(87)90346-7}.

\bibitem{Gross:1987kza}
D.~J. Gross and P.~F. Mende, {The High-Energy Behavior of String Scattering
  Amplitudes}, \emph{Phys. Lett.} {\bf B197}, \penalty0 129,  (1987).
\newblock \doi{10.1016/0370-2693(87)90355-8}.

\bibitem{Gross:1987ar}
D.~J. Gross and P.~F. Mende, {String Theory Beyond the Planck Scale},
  \emph{Nucl. Phys.} {\bf B303}, \penalty0 407,  (1988).
\newblock \doi{10.1016/0550-3213(88)90390-2}.

\bibitem{Amati:1987uf}
D.~Amati, M.~Ciafaloni, and G.~Veneziano, {Classical and Quantum Gravity
  Effects from Planckian Energy Superstring Collisions}, \emph{Int. J. Mod.
  Phys.} {\bf A3}, \penalty0 1615--1661,  (1988).
\newblock \doi{10.1142/S0217751X88000710}.

\bibitem{Amati:1988tn}
D.~Amati, M.~Ciafaloni, and G.~Veneziano, {Can Space-Time Be Probed Below the
  String Size?}, \emph{Phys. Lett.} {\bf B216}, \penalty0 41,  (1989).
\newblock \doi{10.1016/0370-2693(89)91366-X}.

\bibitem{Moeller:2005ez}
N.~Moeller and P.~C. West, {Arbitrary four string scattering at high energy and
  fixed angle}, \emph{Nucl. Phys.} {\bf B729}, \penalty0 1--48,  (2005).
\newblock \doi{10.1016/j.nuclphysb.2005.09.036}.

\bibitem{Giddings:2007bw}
S.~B. Giddings, D.~J. Gross, and A.~Maharana, {Gravitational effects in
  ultrahigh-energy string scattering}, \emph{Phys. Rev.} {\bf D77}, \penalty0
  046001,  (2008).
\newblock \doi{10.1103/PhysRevD.77.046001}.

\bibitem{Taronna:2010qq}
M.~Taronna.
\newblock \emph{{Higher Spins and String Interactions}}.
\newblock PhD thesis, Pisa U.,  (2010).
\newblock URL
  \url{http://inspirehep.net/record/855395/files/arXiv:1005.3061.pdf}.

\bibitem{Sagnotti:2010at}
A.~Sagnotti and M.~Taronna, {String Lessons for Higher-Spin Interactions},
  \emph{Nucl. Phys.} {\bf B842}, \penalty0 299--361,  (2011).
\newblock \doi{10.1016/j.nuclphysb.2010.08.019}.

\bibitem{Fotopoulos:2010ay}
A.~Fotopoulos and M.~Tsulaia, {On the Tensionless Limit of String theory, Off -
  Shell Higher Spin Interaction Vertices and BCFW Recursion Relations},
  \emph{JHEP}. {\bf 11}, \penalty0 086,  (2010).
\newblock \doi{10.1007/JHEP11(2010)086}.

\bibitem{Polyakov:2010sk}
D.~Polyakov, {Higher Spins and Open Strings: Quartic Interactions}, \emph{Phys.
  Rev.} {\bf D83}, \penalty0 046005,  (2011).
\newblock \doi{10.1103/PhysRevD.83.046005}.

\bibitem{Polyakov:2015usr}
D.~Polyakov, {Higher Spins at the Quintic Order: Localization Effect and
  Simplifications}, \emph{Phys. Rev.} {\bf D93}\penalty0 (4), \penalty0 045001,
   (2016).
\newblock \doi{10.1103/PhysRevD.93.045001}.

\bibitem{Sezgin:2002rt}
E.~Sezgin and P.~Sundell, {Massless higher spins and holography}, \emph{Nucl.
  Phys.} {\bf B644}, \penalty0 303--370,  (2002).
\newblock \doi{10.1016/S0550-3213(02)00739-3}.
\newblock [Erratum: Nucl. Phys.B660,403(2003)].

\bibitem{Klebanov:2002ja}
I.~R. Klebanov and A.~M. Polyakov, {AdS dual of the critical O(N) vector
  model}, \emph{Phys. Lett.} {\bf B550}, \penalty0 213--219,  (2002).
\newblock \doi{10.1016/S0370-2693(02)02980-5}.

\bibitem{Boulanger:2008tg}
N.~Boulanger, S.~Leclercq, and P.~Sundell, {On The Uniqueness of Minimal
  Coupling in Higher-Spin Gauge Theory}, \emph{JHEP}. {\bf 08}, \penalty0 056,
  (2008).
\newblock \doi{10.1088/1126-6708/2008/08/056}.

\bibitem{Kessel:2015kna}
P.~Kessel, G.~Lucena~Gómez, E.~Skvortsov, and M.~Taronna, {Higher Spins and
  Matter Interacting in Dimension Three}, \emph{JHEP}. {\bf 11}, \penalty0 104,
   (2015).
\newblock \doi{10.1007/JHEP11(2015)104}.

\bibitem{Boulanger:2015ova}
N.~Boulanger, P.~Kessel, E.~D. Skvortsov, and M.~Taronna, {Higher Spin
  Interactions in Four Dimensions: Vasiliev vs. Fronsdal}.  (2015).

\bibitem{Metsaev:1991mt}
R.~R. Metsaev, {Poincare invariant dynamics of massless higher spins: Fourth
  order analysis on mass shell}, \emph{Mod. Phys. Lett.} {\bf A6}, \penalty0
  359--367,  (1991).
\newblock \doi{10.1142/S0217732391000348}.

\bibitem{Metsaev:2005ar}
R.~R. Metsaev, {Cubic interaction vertices of massive and massless higher spin
  fields}, \emph{Nucl. Phys.} {\bf B759}, \penalty0 147--201,  (2006).
\newblock \doi{10.1016/j.nuclphysb.2006.10.002}.

\bibitem{Vasilev:2011xf}
M.~A. Vasiliev, {Cubic Vertices for Symmetric Higher-Spin Gauge Fields in
  $(A)dS_d$}, \emph{Nucl. Phys.} {\bf B862}, \penalty0 341--408,  (2012).
\newblock \doi{10.1016/j.nuclphysb.2012.04.012}.

\bibitem{Joung:2011ww}
E.~Joung and M.~Taronna, {Cubic interactions of massless higher spins in (A)dS:
  metric-like approach}, \emph{Nucl. Phys.} {\bf B861}, \penalty0 145--174,
  (2012).
\newblock \doi{10.1016/j.nuclphysb.2012.03.013}.

\bibitem{Boulanger:2012dx}
N.~Boulanger, D.~Ponomarev, and E.~D. Skvortsov, {Non-abelian cubic vertices
  for higher-spin fields in anti-de Sitter space}, \emph{JHEP}. {\bf 05},
  \penalty0 008,  (2013).
\newblock \doi{10.1007/JHEP05(2013)008}.

\bibitem{Boulanger:2013zza}
N.~Boulanger, D.~Ponomarev, E.~D. Skvortsov, and M.~Taronna, {On the uniqueness
  of higher-spin symmetries in AdS and CFT}, \emph{Int. J. Mod. Phys.} {\bf
  A28}, \penalty0 1350162,  (2013).
\newblock \doi{10.1142/S0217751X13501625}.

\bibitem{Joung:2015eny}
E.~Joung, S.~Nakach, and A.~A. Tseytlin, {Scalar scattering via conformal
  higher spin exchange}.  (2015).

\bibitem{Vasiliev:2015wma}
M.~Vasiliev, {Star-Product Functions in Higher-Spin Theory and Locality}.
  (2015).

\bibitem{Skvortsov:2015lja}
E.~D. Skvortsov and M.~Taronna, {On Locality, Holography and Unfolding},
  \emph{JHEP}. {\bf 11}, \penalty0 044,  (2015).
\newblock \doi{10.1007/JHEP11(2015)044}.

\bibitem{Krasil'shchik:2010ij}
J.~Krasil'shchik and A.~Verbovetsky, {Geometry of jet spaces and integrable
  systems}, \emph{J. Geom. Phys.} {\bf 61}, \penalty0 1633--1674,  (2011).
\newblock \doi{10.1016/j.geomphys.2010.10.012}.

\bibitem{Barnich:2004cr}
G.~Barnich, M.~Grigoriev, A.~Semikhatov, and I.~Tipunin, {Parent field theory
  and unfolding in BRST first-quantized terms}, \emph{Commun. Math. Phys.} {\bf
  260}, \penalty0 147--181,  (2005).
\newblock \doi{10.1007/s00220-005-1408-4}.

\bibitem{Barnich:2010sw}
G.~Barnich and M.~Grigoriev, {First order parent formulation for generic gauge
  field theories}, \emph{JHEP}. {\bf 01}, \penalty0 122,  (2011).
\newblock \doi{10.1007/JHEP01(2011)122}.

\bibitem{Grigoriev:2012xg}
M.~Grigoriev, {Parent formulations, frame-like Lagrangians, and generalized
  auxiliary fields}, \emph{JHEP}. {\bf 12}, \penalty0 048,  (2012).
\newblock \doi{10.1007/JHEP12(2012)048}.

\bibitem{Rahman:2015pzl}
R.~Rahman and M.~Taronna, {From Higher Spins to Strings: A Primer}.  (2015).

\bibitem{Lopatin:1987hz}
V.~E. Lopatin and M.~A. Vasiliev, {Free Massless Bosonic Fields of Arbitrary
  Spin in $d$-dimensional De Sitter Space}, \emph{Mod. Phys. Lett.} {\bf A3},
  \penalty0 257,  (1988).
\newblock \doi{10.1142/S0217732388000313}.

\bibitem{Vasiliev:1988sa}
M.~A. Vasiliev, {Consistent Equations for Interacting Massless Fields of All
  Spins in the First Order in Curvatures}, \emph{Annals Phys.} {\bf 190},
  \penalty0 59--106,  (1989).
\newblock \doi{10.1016/0003-4916(89)90261-3}.

\bibitem{Vasiliev:1999ba}
M.~A. Vasiliev, {Higher spin gauge theories: Star product and AdS space}.
  (1999).

\bibitem{Joung:2013doa}
E.~Joung, M.~Taronna, and A.~Waldron, {A Calculus for Higher Spin
  Interactions}, \emph{JHEP}. {\bf 07}, \penalty0 186,  (2013).
\newblock \doi{10.1007/JHEP07(2013)186}.

\bibitem{Prokushkin:1999xq}
S.~F. Prokushkin and M.~A. Vasiliev, {Cohomology of arbitrary spin currents in
  AdS(3)}, \emph{Theor. Math. Phys.} {\bf 123}, \penalty0 415--435,  (2000).
\newblock \doi{10.1007/BF02551048}.
\newblock [Teor. Mat. Fiz.123,3(2000)].

\bibitem{Bekaert:2014cea}
X.~Bekaert, J.~Erdmenger, D.~Ponomarev, and C.~Sleight, {Towards holographic
  higher-spin interactions: Four-point functions and higher-spin exchange},
  \emph{JHEP}. {\bf 03}, \penalty0 170,  (2015).
\newblock \doi{10.1007/JHEP03(2015)170}.

\bibitem{Bekaert:2015tva}
X.~Bekaert, J.~Erdmenger, D.~Ponomarev, and C.~Sleight, {Quartic AdS
  Interactions in Higher-Spin Gravity from Conformal Field Theory},
  \emph{JHEP}. {\bf 11}, \penalty0 149,  (2015).
\newblock \doi{10.1007/JHEP11(2015)149}.

\bibitem{Sciuto:1969vz}
S.~Sciuto, {The general vertex function in dual resonance models}, \emph{Lett.
  Nuovo Cim.} {\bf 2S1}, \penalty0 411--418,  (1969).
\newblock \doi{10.1007/BF02755622}.

\bibitem{DiVecchia:1986mb}
P.~Di~Vecchia, R.~Nakayama, J.~L. Petersen, J.~R. Sidenius, and S.~Sciuto,
  {Covariant N String Amplitude}, \emph{Nucl. Phys.} {\bf B287}, \penalty0 621,
   (1987).
\newblock \doi{10.1016/0550-3213(87)90121-0}.

\bibitem{Neveu:1986ai}
A.~Neveu and P.~C. West, {The Cyclic Symmetric Vertex for Three Arbitrary
  Neveu-schwarz Strings}, \emph{Phys. Lett.} {\bf B180}, \penalty0 34,  (1986).
\newblock \doi{10.1016/0370-2693(86)90129-2}.

\bibitem{Gross:1986ia}
D.~J. Gross and A.~Jevicki, {Operator Formulation of Interacting String Field
  Theory}, \emph{Nucl. Phys.} {\bf B283}, \penalty0 1,  (1987).
\newblock \doi{10.1016/0550-3213(87)90260-4}.

\bibitem{DiVecchia:1988hq}
P.~Di~Vecchia, K.~Hornfeck, M.~Frau, A.~Lerda, and S.~Sciuto, {$N$ String, $g$
  Loop Vertex for the Bosonic String}, \emph{Phys. Lett.} {\bf B206}, \penalty0
  643--649,  (1988).
\newblock \doi{10.1016/0370-2693(88)90711-3}.

\bibitem{Gelfond:2006be}
O.~A. Gelfond, E.~D. Skvortsov, and M.~A. Vasiliev, {Higher spin conformal
  currents in Minkowski space}, \emph{Theor. Math. Phys.} {\bf 154}, \penalty0
  294--302,  (2008).
\newblock \doi{10.1007/s11232-008-0027-6}.

\bibitem{Gelfond:2010pm}
O.~A. Gelfond and M.~A. Vasiliev, {Unfolded Equations for Current Interactions
  of 4d Massless Fields as a Free System in Mixed Dimensions}, \emph{J. Exp.
  Theor. Phys.} {\bf 120}\penalty0 (3), \penalty0 484--508,  (2015).
\newblock \doi{10.1134/S106377611503005X}.

\bibitem{Gelfond:2014pja}
O.~A. Gelfond and M.~A. Vasiliev, {Conserved Higher-Spin Charges in $AdS_4$}.
  (2014).

\bibitem{Chang:2011mz}
C.-M. Chang and X.~Yin, {Higher Spin Gravity with Matter in $AdS_3$ and Its CFT
  Dual}, \emph{JHEP}. {\bf 10}, \penalty0 024,  (2012).
\newblock \doi{10.1007/JHEP10(2012)024}.

\bibitem{Chang:2011vka}
C.-M. Chang and X.~Yin, {Correlators in $W_N$ Minimal Model Revisited},
  \emph{JHEP}. {\bf 10}, \penalty0 050,  (2012).
\newblock \doi{10.1007/JHEP10(2012)050}.

\bibitem{Ammon:2011ua}
M.~Ammon, P.~Kraus, and E.~Perlmutter, {Scalar fields and three-point functions
  in $D=3$ higher spin gravity}, \emph{JHEP}. {\bf 07}, \penalty0 113,  (2012).
\newblock \doi{10.1007/JHEP07(2012)113}.

\bibitem{Fitzpatrick:2014vua}
A.~L. Fitzpatrick, J.~Kaplan, and M.~T. Walters, {Universality of Long-Distance
  AdS Physics from the CFT Bootstrap}, \emph{JHEP}. {\bf 08}, \penalty0 145,
  (2014).
\newblock \doi{10.1007/JHEP08(2014)145}.

\bibitem{Alkalaev:2015wia}
K.~B. Alkalaev and V.~A. Belavin, {Classical conformal blocks via AdS/CFT
  correspondence}, \emph{JHEP}. {\bf 08}, \penalty0 049,  (2015).
\newblock \doi{10.1007/JHEP08(2015)049}.

\bibitem{Taronna:2011kt}
M.~Taronna, {Higher-Spin Interactions: four-point functions and beyond},
  \emph{JHEP}. {\bf 04}, \penalty0 029,  (2012).
\newblock \doi{10.1007/JHEP04(2012)029}.

\bibitem{Taronna:2012gb}
M.~Taronna.
\newblock \emph{{Higher-Spin Interactions: three-point functions and beyond}}.
\newblock PhD thesis, Pisa, Scuola Normale Superiore,  (2012).
\newblock URL
  \url{https://inspirehep.net/record/1188191/files/arXiv:1209.5755.pdf}.

\bibitem{Polchinski:1999ry}
J.~Polchinski, {S matrices from AdS space-time}.  (1999).

\bibitem{Giddings:1999jq}
S.~B. Giddings, {Flat space scattering and bulk locality in the AdS / CFT
  correspondence}, \emph{Phys. Rev.} {\bf D61}, \penalty0 106008,  (2000).
\newblock \doi{10.1103/PhysRevD.61.106008}.

\bibitem{Gary:2009ae}
M.~Gary, S.~B. Giddings, and J.~Penedones, {Local bulk S-matrix elements and
  CFT singularities}, \emph{Phys. Rev.} {\bf D80}, \penalty0 085005,  (2009).
\newblock \doi{10.1103/PhysRevD.80.085005}.

\bibitem{Fitzpatrick:2012yx}
A.~L. Fitzpatrick, J.~Kaplan, D.~Poland, and D.~Simmons-Duffin, {The Analytic
  Bootstrap and AdS Superhorizon Locality}, \emph{JHEP}. {\bf 12}, \penalty0
  004,  (2013).
\newblock \doi{10.1007/JHEP12(2013)004}.

\bibitem{Mack:2009mi}
G.~Mack, {D-independent representation of Conformal Field Theories in D
  dimensions via transformation to auxiliary Dual Resonance Models. Scalar
  amplitudes}.  (2009).

\bibitem{Penedones:2010ue}
J.~Penedones, {Writing CFT correlation functions as AdS scattering amplitudes},
  \emph{JHEP}. {\bf 03}, \penalty0 025,  (2011).
\newblock \doi{10.1007/JHEP03(2011)025}.

\bibitem{Fitzpatrick:2011ia}
A.~L. Fitzpatrick, J.~Kaplan, J.~Penedones, S.~Raju, and B.~C. van Rees, {A
  Natural Language for AdS/CFT Correlators}, \emph{JHEP}. {\bf 11}, \penalty0
  095,  (2011).
\newblock \doi{10.1007/JHEP11(2011)095}.

\bibitem{Paulos:2011ie}
M.~F. Paulos, {Towards Feynman rules for Mellin amplitudes}, \emph{JHEP}. {\bf
  10}, \penalty0 074,  (2011).
\newblock \doi{10.1007/JHEP10(2011)074}.

\bibitem{Fitzpatrick:2011hu}
A.~L. Fitzpatrick and J.~Kaplan, {Analyticity and the Holographic S-Matrix},
  \emph{JHEP}. {\bf 10}, \penalty0 127,  (2012).
\newblock \doi{10.1007/JHEP10(2012)127}.

\bibitem{Fitzpatrick:2011dm}
A.~L. Fitzpatrick and J.~Kaplan, {Unitarity and the Holographic S-Matrix},
  \emph{JHEP}. {\bf 10}, \penalty0 032,  (2012).
\newblock \doi{10.1007/JHEP10(2012)032}.

\bibitem{Dempster:2012vw}
P.~Dempster and M.~Tsulaia, {On the Structure of Quartic Vertices for Massless
  Higher Spin Fields on Minkowski Background}, \emph{Nucl. Phys.} {\bf B865},
  \penalty0 353--375,  (2012).
\newblock \doi{10.1016/j.nuclphysb.2012.07.031}.

\end{thebibliography}

\end{document}